\documentclass[epj,final]{svjour}

\usepackage{color}
\usepackage{amsmath}
\usepackage{amssymb}
\usepackage{graphicx,epsfig}
\usepackage{bm}

\newcommand{\be}{\begin{equation}}
\newcommand{\ee}{\end{equation}}

\begin{document}

\title{Counter-flow instability of a quantum mixture of two superfluids}
\titlerunning{Counter-flow instability of a quantum mixture of two superfluids}

\author{Marta Abad\inst{1} \and Alessio Recati\inst{1} \and Sandro Stringari\inst{1} \and Fr\'ed\'eric Chevy\inst{2}}
\institute{INO-CNR BEC Center and Dipartimento di Fisica, Universit\`a di Trento, I-38123 Povo, Italy \and Laboratoire Kastler-Brossel, \'Ecole Normale Sup\'erieure, CNRS and UPMC, 24 rue Lhomond, 75005 Paris, France}

\abstract{
We study the instability of a mixture of two interacting counter-flowing superfluids. For a homogeneous system, we show that superfluid hydrodynamics leads to the existence of a dynamical instability at a critical value of the relative velocity $v_{cr}$.  When the interspecies coupling is small the critical value approaches the value $v_{cr}=c_1+c_2$, given by the sum of  the sound velocities of the two uncoupled superfluids, in agreement with the recent prediction of \cite{castin2014landau} based on Landau's argument.  The crucial dependence of the critical velocity on the interspecies coupling is explicitly discussed. Our results agree with previous predictions for weakly interacting Bose-Bose mixtures and applies to Bose-Fermi superfluid mixtures as well.  Results for the stability of transversally trapped mixtures are also presented.
}



\maketitle

Recent years have witnessed the advent of ultracold gases as a unique playground for the study of quantum many-body phenomena \cite{bloch2008many}. The wide tunability of these systems allowed experimentalists to engineer a broad range of physical situations and paved the way to the experimental study of outstanding problems in condensed matter physics or astrophysics.

An important direction of research concerns the study of quantum mixtures of two
superfluids and the onset of their instability when they move against each other (counterflow
instability). 
In the case of weakly interacting Bose-Einstein condensates the dynamical counterflow instability has been theoretically studied in several works \cite{law2001critical,Yukalov2004,Kravchenko2008,Kravchenko2009,Takeuchi2010,Ishino2011} and has been experimentally seen to lead to the formation of soliton trains \cite{hamner2011generation,hoefer2011}. 
The recent results presented in \cite{Ferrier2014Mixture} for mixtures of Bosons and Fermions demonstrated the possibility of reaching double superfluidity with atomic  gases belonging to different statistics, opening a new scenario for the physics of superfluid mixtures. The study of the collective excitations of the mixture raised, in particular, the question of the critical velocity of their relative superfluid motion. 

In \cite{castin2014landau}, Castin {\em et al.} proposed a generalization of Landau's mechanism applied to a mixture with vanishingly small interspecies coupling, where the relative flow decays by shedding pairs of elementary excitations in the two components of the mixture. When the excitation spectra of the two superfluids are dominated by acoustic modes, the critical velocity $v_{cr}$ is the sum of their sound velocities:  $v_{cr}= c_1 +c_2$. A similar result was obtained in \cite{zheng2014quasi} by considering the lifetime of the quasi-particles of the system. This result for the critical
velocity   differs from the usual Landau's prescription $v_{cr}=\text{min}(c_1,c_2)$ holding for two independent superfluids. Actually the latter result holds only if the superfluids can exchange momentum with an external wall or with a moving impurity. Instead, in \cite{castin2014landau}, momentum conservation is ensured by the excitation of two phonons with opposite momenta. 

In the present article, we study the stability of a superfluid counter-flow in the hydrodynamic approximation, discussing in an explicit way the role of the interspecies interaction, going beyond the small interspecies coupling limit discussed in \cite{castin2014landau}.

Consider a mixture of two superfluids labeled by the indices $\alpha=1,2$.  In the hydrodynamic approximation, the equations of motion at zero temperature can be obtained starting from the energy functional
\begin{align}
&E[n_{1},n_{2},\phi_{1},\phi_{2}]= \int\bigg[ \frac{1}{2}m_{1}n_{1}(\nabla\phi_{1})^2 + \frac{1}{2}m_{2}n_{2}(\nabla\phi_{2})^2 + \nonumber\\
&+ U_{1}n_{1} + U_{2}n_{2}+e_{11}(n_{1}) + e_{22}(n_{2}) + e_{12}(n_{1},n_{2})  \bigg] \text{d}\bm{r} \label{EqEF}
\end{align}
where $m_\alpha$ is the mass of the particles of the superfluid $\alpha=1,2$ and $U_\alpha(\bm{r})$ their trapping potential. The terms $e_{11}(n_{1})$, $e_{2}(n_{2})$ and $e_{12}(n_{1},n_{2})$ are the energy densities corresponding, respectively, to the interactions among particles $1$, particles $2$ and between particles $1$ and $2$. The velocity potential, $\phi_{\alpha}$, is related to the superfluid velocity of each fluid as $\bm{v}_{\alpha}=\nabla \phi_{\alpha}$.
Choice (\ref{EqEF}) for the energy functional ignores the possible coupling between the velocity fields of the two fluids, which lead to physical phenomena such as the Andreev-Bashkin effect~\cite{AndreevBashkin}. This effect is however expected to be small in dilute gases.

The equations for the density and velocity fields, $n_\alpha (\bm r,t)$ and $\bm v_\alpha(\bm r,t)$, can be derived by taking the velocity potential and the density as conjugate variables obeying Hamilton's equations,
$\partial n_\alpha/\partial t= \delta E/ \delta \phi_\alpha$ and $\partial \phi_\alpha /\partial t= - \delta E/ \delta
 n_\alpha$. They take the form
\begin{eqnarray}
m_\alpha\left(\partial_t\bm v_\alpha+\bm\nabla v_\alpha^2/2\right)&=&-\bm\nabla\left(U_\alpha+\mu_\alpha\right)\\
\partial_t n_\alpha+\bm\nabla(n_\alpha \bm v_\alpha)&=&0,
\end{eqnarray}
where $\mu_\alpha(n_1,n_2)$ is the chemical potential of each species at rest, given by
\begin{eqnarray}
\mu_1(n_1,n_2)&=&\frac{\partial e_{11}(n_1)}{\partial n_1}  +\frac{\partial e_{12}(n_1,n_2)}{\partial n_1}
\label{Eq:ChemicalPotential1}\\
\mu_2(n_1,n_2)&=&\frac{\partial e_{22}(n_2)}{\partial n_2}  +\frac{\partial e_{12}(n_1,n_2)}{\partial n_2}
\label{Eq:ChemicalPotential2}
\end{eqnarray}

We address first the case of a homogeneous system ($U_\alpha=0$). The stationary solutions correspond to uniform densities and velocity fields, with $n_\alpha=n_{\alpha}^{(0)}$ and $\bm v_\alpha=\bm v_{\alpha}^{(0)}$.  We consider the general case $\bm v_1^{(0)} \ne \bm v_2^{(0)}$ corresponding to a non vanishing counter-flow velocity $\bm v=\bm v_1^{(0)}-\bm v_2^{(0)}$.  If the system is weakly perturbed with respect to the stationary configuration, we can look for solutions in the linear regime:
\begin{eqnarray}
n_\alpha(\bm r,t)=n_\alpha^{(0)}+n_\alpha^{(1)}e^{i(\bm q\cdot \bm r-\omega t)}\\
\bm v_\alpha(\bm r,t)=\bm v_\alpha^{(0)}+\bm w_\alpha e^{i(\bm q\cdot \bm r-\omega t)}.
\end{eqnarray}
Expanding the hydrodynamic equations to first order in the perturbation yields
\begin{eqnarray}
-im_\alpha\left(\omega- q\bm v_\alpha^{(0)}\cdot\bm u\right)\bm w_\alpha&=&-i\bm q\sum_\beta \frac{\partial\mu_\alpha}{\partial n_\beta} n_\beta^{(1)}\label{Eq:Linearized1}\\
i\left(\omega- q\bm v_\alpha^{(0)}\cdot\bm u\right) n_\alpha^{(1)}&=&i\bm q\left(n_\alpha^{(0)} \bm w_\alpha\right).\label{Eq:Linearized2}
\end{eqnarray}
with $\bm u=\bm q/q$ a unitary vector in the direction of the quasimomentum $\bm q$. 
Using Eq. (\ref{Eq:Linearized1}) to eliminate $\bm w_\alpha$ from Eq. (\ref{Eq:Linearized2}) we obtain
\be
m_\alpha (\omega- q \bm v_\alpha^{(0)}\cdot\bm u)^2n_\alpha^{(1)}=q^2 n_\alpha^{(0)}\sum_{\beta}\frac{\partial\mu_\alpha}{\partial n_\beta}n_\beta^{(1)}
\label{Eq6}
\ee
By writing  the frequency $\omega$ of the solution as $\omega=c q$, we find that the sound velocity $c$ should satisfy the condition
\be
\left[(c- \bm v_1^{(0)}\cdot\bm u)^2-c_1^2\right]\left[(c- \bm v_2^{(0)}\cdot\bm u)^2-c_2^2\right]=c^4_{12}\,.
\label{Eq7}\ee
In the above equation we have introduced the quantities
\begin{align}
 &c_\alpha^{2} = \frac{n_{\alpha}^{(0)}}{m_{\alpha}}\frac{\partial \mu_{\alpha}}{\partial n_{\alpha}} \\
 &c^2_{12}=\sqrt{\frac{n_1^{(0)}n_2^{(0)}}{m_1m_2}\frac{\partial \mu_1}{\partial n_2}\frac{\partial \mu_2}{\partial n_1}}
\end{align}
Notice that the hydrodynamic formalism is valid in the long wavelength limit, that is when $q\to0$, and for sound-like excitations. In the following we will always consider that the system is dynamically stable at rest, that is $c_{12}^2<c_1c_2$.

When the intercomponent interaction can be treated within the mean-field approximation, we have 
\begin{eqnarray}
\mu_1(n_1,n_2)=\mu_1^{(0)}(n_1)+g_{12}n_2\\
\mu_2(n_1,n_2)=\mu_2^{(0)}(n_2)+g_{12}n_1,
\end{eqnarray}
where $\mu_\alpha^{(0)}$ is the chemical potential of species $\alpha$ alone and $g_{12}$ is the coupling constant between the two species. In this case, the quantities $c_1$ and $c_2$ coincide with the sound velocities of the two uncoupled fluids at rest and the cross-term $c_{12}^2$ is proportional to $g_{12}$.

The solutions of Eq.~(\ref{Eq7}) are not always real and can therefore lead to a dynamic instability of the system. Owing to Galilean Invariance, the instability rate depends only on the relative velocity of the two fluids. Indeed, let's consider a Galilean boost of the two superfluids $\bm v_\alpha^{(0)}\rightarrow \bm v_\alpha^{(0)}+\bm V$. The solution of Eq.~(\ref{Eq7}) is simply shifted by the same amount and we have $c\rightarrow c+\bm V\cdot\bm u$. In particular, its imaginary part is not modified by the boost. From this property we see that for any $\bm V$, ${\rm Im} [c(\bm v_1^{(0)}+\bm V,\bm v_2^{(0)}+\bm V)]={\rm Im}[c(\bm v_1^{(0)},\bm v_2^{(0)})]$, where ${\rm Im}$ is the imaginary part. Taking $\bm V=-\bm v_2^{(0)}$, we deduce that ${\rm Im} [c(\bm v_1^{(0)},\bm v_2^{(0)})]={\rm Im} [c(\bm v_1^{(0)}-\bm v_2^{(0)},0)]$ and, as expected, depends only on the relative velocity $\bm v$.

\begin{figure}
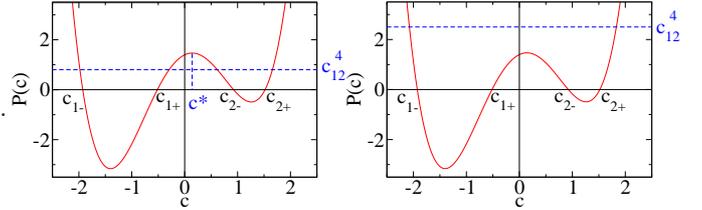

\epsfig{file=pola.eps, width=0.5\linewidth, clip=true}%
\epsfig{file=polb.eps, width=0.5\linewidth, clip=true}
\caption{Left: Typical graph of $P(c)$. $c_{i,\pm}$ are the roots of $P$ and $c^*$ is the abscissa of the maximum of $P$. For $c^4_{12} < P(c^*)$ there are four real solutions of  $P(c)=c^4_{12}$, hence no imaginary solution. Right: For $c^4_{12}> P(c^*)$, then the equation $P(c)=c^4_{12}$ has  two complex roots, leading to an instability. The instability threshold condition is therefore given by  $P(c^*)=c^4_{12}$.}
\label{Fig1}
\end{figure}

In order to discuss the  general solutions of Eq.~(\ref{Eq7})  it is convenient to write it as
\be
P(c)=c^4_{12}
\label{Eq8}
\ee
where $P(c)=\left[\left(c-\bm v_1^{(0)}\cdot\bm u\right)^2-c_1^2\right]\left[\left(c- \bm v_2^{(0)}\cdot\bm u\right)^2-c_2^2\right]$ is a fourth-degree polynomial. 
$P(c)$ has four roots, $c_{i\pm}= \bm v_i^{(0)}\cdot\bm u\pm c_i$ and its typical shape is displayed in Fig. \ref{Fig1}  where $c^*$ corresponds to the maximum of $P$. If $c^4_{12}$  is below $P(c^*)$, then Eq. (\ref{Eq8}) has four real roots and the mixture is dynamically stable. If instead  $c^4_{12}$ is above $P(c^*)$, then Eq. (\ref{Eq8}) has 2 real roots, and 2  complex conjugate roots, leading to a dynamical instability of the system. The instability threshold thus corresponds to the condition $P(c^*)=c^4_{12}$.

We first note that the system is always unstable when $c_{12}^{2}>c_1 c_2$. Indeed, consider a wave-vector $\bm q$ orthogonal to both $\bm v_\alpha^{(0)}$. In this case, $P(c)=(c^2-c_1^2)(c^2-c_2^2)$ and its maximum is achieved for $c^*=0$ with $P(c^*)=c_1^2c_2^2$. 
We therefore obtain the well-known instability criterion for the phase separation of a static mixture of two superfluids.

Let us now consider the situation of a  small coupling  $c^2_{12}$. The condition $P(c^*)=c^4_{12}\simeq 0$ implies that $c^*$ is a root of both $P$ and $P'$ and is thus a double root of $P$. The only way to satisfy this condition is to have $c_{1\pm}\simeq c_{2\mp}\simeq c^*$.
Close  to the onset of instability, we can approximate $P(c)$ in the vicinity of $c^*$ by
\be
P(c)\simeq -4 c_1c_2(c-c_{1+})(c-c_{2-}).
\ee
Since the extremum of a parabola is at the middle of its roots, we have $c^*=(c_{1+}+c_{2-})/2$ and  hence
\be
P(c^*)\simeq c_1c_2(c_{2-}-c_{1+})^2\simeq c_1c_2(\bm v\cdot\bm u-c_1-c_2)^2.
\ee
Since the mixture is unstable when $P(c^*)\le c_{12}^4$, we see that the width of the instability domain is given by the inequality
\be
\left|\bm v\cdot\bm u- c_1-c_2\right|\le \frac{c^2_{12}}{\sqrt{c_1c_2}} \;, \label{Eq:Delta0}
\ee
i.e. takes place in the interval $c_1+c_2 - c^2_{12}/\sqrt{c_1c_2}< |\bm v\cdot\bm u| <c_1+c_2 +c^2_{12}/\sqrt{c_1c_2}$.
Note that in the unstable region, the imaginary part of $c$ (yielding the departure rate from equilibrium) is given by
\be
{\rm Im}(c)=\frac{1}{2}\sqrt{\frac{c^4_{12}}{c_1c_2} -(c_{1+}-c_{2-})^2}\le \frac{1}{2}\frac{c^2_{12}}{\sqrt{c_1c_2}}\,,
\label{Eq9}
\ee
the upper-bound  being  achieved when $c_{1+}=c_{2-}$, i.e. when $|\bm v\cdot\bm u|=c_1+c_2$. Since $|\bm v\cdot\bm u|\le v$, we see that the counter flow is unstable as soon as $v\ge v_{cr}=c_1+c_2$, which corresponds to the Landau criterion predicted in \cite{castin2014landau}. Above this threshold the counter-flow decays by shedding excitations of opposite momenta propagating along a direction $\bm u$ such that $|\bm v\cdot\bm u|=v_{cr}$.

Notice finally that the  imaginary part of $c$ becomes smaller and smaller  as the interspecies coupling $c^2_{12}$ tends to zero. This implies that the instability rate goes to zero for vanishingly small couplings, even though the critical velocity remains finite.

In the experiment described in \cite{Ferrier2014Mixture}, the sound velocity of the Bose gas is about five times smaller than that of the unitary Fermi superfluid. It is therefore interesting to calculate the critical velocity when one of the sound velocities (for example $c_{2}$) tends to zero.
To this purpose, it is convenient to expand the discriminant of the polynomial $P(c)-c^{4}_{12}$ to the lowest nonzero order in $c_{2}$ (second order) and equate it to zero, as was sketched in~\cite{law2001critical}. This leads to the values of $v$ at which $P(c^{*})=c_{12}^{4}$, yielding the result
\begin{equation}
v_{\text{cr}} = c_1\sqrt{1-\Delta^{2}}
\label{vcrcb=0}
\end{equation}
for the onset of dynamic instability, where we have introduced the dimensionless parameter $\Delta=\frac{c_{12}^2}{c_{1}c_{2}}$.

Having understood the behavior of the solutions of the hydrodynamic equations for small values of the interspecies coupling and $c_{2}$ we can now investigate the effect for arbitrary values of $c_{2}$ and $c_{12}$ by solving numerically the relevant Eq. (\ref{Eq7}). To this purpose it is convenient to rewrite (\ref{Eq7}) in dimensionless units, dividing it by $c_{1}^{2}c_{2}^{2}$ and expressing all the velocities in units of $c_1$: $c \to \tilde{c}=c/c_1$, $c_2 \to \tilde{c}_{2}=c_{2}/c_1$, $\bm v_\alpha^{(0)}\to \tilde{\bm v}_\alpha^{(0)}=\bm v_\alpha^{(0)}/c_1$.
The solutions for $\tilde{c}$ then depend on $\tilde{\bm v}_\alpha^{(0)}$, $\tilde{c_{2}}$ and $\Delta$. As justified above, the onset of dynamical instability depends on $\tilde{\bm v}=\tilde{\bm v}_1^{(0)}-\tilde{\bm v}_2^{(0)}$ as well as $\tilde{c_{2}}$ and $\Delta$.

\begin{figure}
	\epsfig{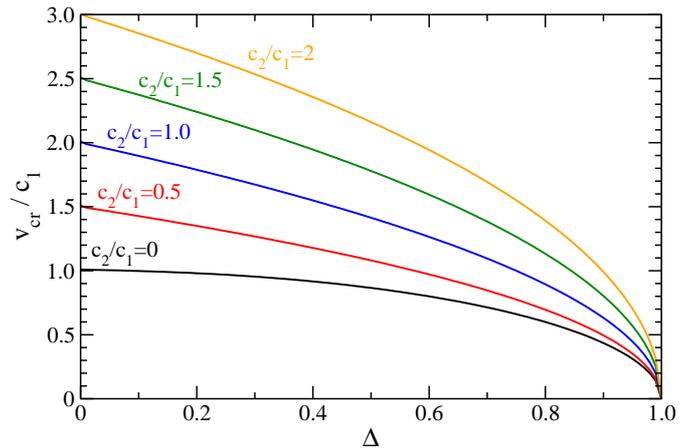}
	\caption{Critical relative velocity, $v/c_{1}$, as a function of the interfluid coupling $\Delta$, for various values of $c_{2}/c_{1}$.}\label{Figvcr}
\end{figure}

In Fig.~\ref{Figvcr} we show the results for the critical velocity, $\tilde{v}_{cr}$, associated with the emergence  of a complex solution of $\tilde{c}$ as a function of $\Delta$, for different choices of the ratio $\tilde{c}_2$. Some comments are in order here.

i) For $\Delta \to 0$ the results are consistent with the prediction $ v_{cr}= c_1 + c_2$ found by Castin {\it et al.} and  discussed above. In general we find that the critical value $v_{cr}$ given by the condition Eq.~(\ref{Eq:Delta0}) for small values of $\Delta$ provides a good approximation to the exact solution up to $\Delta\sim0.3$.

ii) The critical velocity vanishes as $\Delta \to 1$ ($c_{1}c_{2}\to c_{12}^{2}  $), corresponding to the onset of dynamic instability in the absence of moving fluids discussed above.

iii) In addition to the two previous cases, a simple analytical solution can be obtained for $c_{1}=c_{2}\equiv c_{0}$ for which we find
\be
	c(\bm v)=\frac{1}{2}\sqrt{(\bm v\cdot\bm u)^{2}+4c_{0}^{2}\pm4\sqrt{(\bm v\cdot\bm u)^{2}c_{0}^{2}+c_{12}^{4}}}\ .\label{Eq:c1eqc2}
\ee
The condition for dynamical instability is $|(\bm v\cdot \bm u)^{2}-4c_{0}^{2}|=4c_{12}^{2}$, which leads to the critical velocities
\be 
v_{cr}^{\pm}=2c_{0}\sqrt{1\pm\Delta}\,.
\ee
The system is thus dynamically unstable in the range $v_{cr}^{-}<|\bm v\cdot\bm u|<v_{cr}^{+}$, whereas remains dynamically stable outside this region. 

Let us make a remark here on the upper bound of the instability zone, $v_{cr}^+$. This result emerges  naturally in hydrodynamics where the dispersion is linear in $q$. However, inclusion of dispersive terms in the excitation spectrum can give rise to
a different scenario. For instance, in the context of bosonic mixtures, where the excitation spectrum is known for all $q$, it can be shown that for $|\bm v\cdot\bm u|>v_{cr}^{+}$ the dynamical instability arises at finite values of $q$ \cite{law2001critical,Ishino2011}. Therefore even if two superfluids moving at a relative velocity $|\bm v\cdot\bm u|>v_{cr}^+$ are dynamically stable against the excitation of long wavelength phonons, they can be unstable against the excitation of shorter wavelength modes.


Let us finally comment that our results apply to both Bose-Bose and Bose-Fermi mixtures (the case of Fermi-Fermi mixtures will not be discussed here), in the weak as well as in the strong coupling limit. Let us discuss the explicit form for the various quantities entering the calculation of the critical velocity in the two cases:

a) {\bf Weakly interacting Bose-Bose gases}. In this case  the sound velocities are given by the Bogoliubov form $m_\alpha c_\alpha^2 = g_\alpha n_\alpha$, with the coupling constant fixed by the s-wave scattering length according to $g_{\alpha}= 4 \pi \hbar^2 a_{\alpha }/m_\alpha$. The dimensionless interspecies coupling  $\Delta$ is given by the density-independent result $\Delta=g_{12}/\sqrt{g_{1}g_{2}}$, with $g_{12}=2\pi\hbar^2 a_{12}/m_r$, where we have introduced the reduced mass $m_{r}^{-1}=m_{1}^{-1}+m_{2}^{-1}$. 
Our results agree with those previously reported (see \cite{law2001critical,Yukalov2004,Kravchenko2008,Kravchenko2009,Takeuchi2010,Ishino2011,hoefer2011}) in the limit $q\to0$.

b) {\bf Dilute Bose gas interacting with a unitary Fermi gas}. In this case the sound velocity of the Fermi gas (hereafter called $c_1$) is given by $c_{1}= v_{F}\sqrt{\xi/3}$, with $v_{F}$ the Fermi velocity and $\xi$ the Bertsch parameter \cite{giorgini2008theory},
while the constant $\Delta$ takes the density-dependent form
\begin{equation}
	\Delta^2 = \left(\frac{3}{\pi}\right)^{1/3}\frac{(m_{1}+m_{2})^{2}}{\xi m_{1}m_{2}}\frac{a_{12}^2(n_{1}^{0})^{1/3}} {a_{2}}.
\end{equation}
In the case of $^6$Li-$^7$Li superfluid mixtures, recently experimentally implemented in \cite{Ferrier2014Mixture}, the interaction parameters can be tuned in a rather flexible way thanks to the occurrence of various Feshbach resonances, so that these mixtures are excellent candidates to explore in detail the mechanisms of dynamic instability discussed in the present paper.
Let us also notice that the results presented in this work are based on the assumption that the two fluids are miscible at rest. According to Viverit {\it et al.} \cite{Viverit2000}, for positive values of the interspecies coupling constant $g_{12}$, phase separation in a Bose-Fermi mixture can actually occur before the onset $\Delta=1$ of dynamical instability.

Let's now turn to the case of a transversally trapped system where the external potentials $U_\alpha$ depend on the transverse coordinate $\bm\rho = (x,y)$. The stationary density profiles are then given by the Local Density Approximation (LDA) condition $\mu_\alpha (n_1^{(0)}(\bm\rho),n_2^{(0)}(\bm\rho))+U_\alpha(\bm\rho)=\mu_\alpha^{(0)}$.
From the LDA results for the 3D density one can calculate the double
integrated (1D) densities $\bar n_\alpha=\int d^2\bm\rho n_\alpha^{(0)}$ of the two fluids, in terms of
the chemical potentials $\mu_1$ and $\mu_2$. The knowledge of the equations
of state $\mu_1(\bar{n}_1,\bar{n}_2)$ and
$\mu_2(\bar{n}_1,\bar{n}_2)$ then permits to derive the  1D
hydrodynamic equations, following the same procedure employed in the first
part of the paper,  provided the wave vector $q$ of the sound wave is smaller than the
radial size of the mixtures.
 The simplest case is when the coupling term is
vanishingly small. In this case one finds that the system becomes unstable
for $v_{cr}=\bar{c}_{1}+\bar{c}_{2}$, in analogy with the 3D result, where
$m_\alpha\bar{c}^2_\alpha = \partial\mu_{\alpha}/\partial\bar{n}_{\alpha}$ are the sound velocities of the two
independent fluids in the 1D-like configurations.
In the Bose case one finds $\mu_{\text{Bose}}\propto \bar{n}^{1/2}$ and hence $\bar{c}_{\text{Bose}}= 1/\sqrt{2} \,c_{\text{Bose}}$ \cite{Zaremba1998,stringari1998dynamics}. In the unitary Fermi gas one instead finds $\mu_{\text{Fermi}}\propto \bar{n}^{2/5}$ yielding $\bar{c}_{\text{Fermi}}= \sqrt{3/5}\,c_{\text{Fermi}}$ \cite{Capuzzi2006,Hou:2013}. In the above equations $c_{\text{Bose}}$ and $c_{\text{Fermi}}$ are the sound velocities calculated, for a uniform Bose and Fermi gas, respectively, at the central density. In the experiment reported in \cite{Ferrier2014Mixture}, $\Delta\simeq 10^{-1}$. The effects of the interactions are then negligible and the measured critical velocity is indeed very close to the prediction $v_{\rm cr}=\bar{c}_{\text{Fermi}}+\bar{c}_{\text{Bose}}$.
It is also worth noticing that our results apply to any two species Luttinger
liquid, in particular to spin-1/2 Fermi gases confined to one-dimension.
In the latter case the relative velocity breaks the spin-charge separation
of the system and according to our result produces a dynamical instability
if it is large enough.

In conclusion, using a hydrodynamic approach, we have derived explicit results for
the emergence of dynamic instability in a  mixture of two superfluids, by
calculating the critical velocity associated with the relative motion of the two
components of the mixture. Our results hold also for fluids belonging to different
quantum statistics and generalize previous results derived for mixtures of
Bose-Einstein condensates. For relative velocities larger than the critical value
the  solutions of the hydrodynamic equations exhibit an imaginary component in the
sound velocity which is responsible for decay processes. The role of the
interspecies interaction has been explicitly investigated and  shown to decrease the
value of the critical velocity with respect to the prediction of~\cite{castin2014landau}, derived in the
limit of vanishingly small interspecies coupling constant. Special emphasis has been
given to the behavior of Bose-Fermi superfluid  mixtures   where  experimental
measurements of the collective motion have recently become available.



\vspace{2em}
This work has been supported by ERC through the QGBE and ThermoDynaMix grants, by Provincia Autonoma di Trento and by R\'egion Ile de France (IFRAF). A. R. acknowledges support from the Alexander von Humboldt foundation.


\end{document}